\documentclass[final,5p,times,twocolumn]{elsarticle}
\usepackage{amsmath}
\usepackage{amssymb}
\usepackage{amsthm}
\usepackage{graphicx}

\journal{Solid State Communications}

\begin{document}

\begin{frontmatter}

\title{Translation-invariant bipolarons and the problem of high-temperature superconductivity.}

\author[lak]{V.D.~Lakhno\corref{cor1}}
\ead{lak@impb.psn.ru}

\cortext[cor1]{Corresponding author}

\address[lak]{Institute of Mathematical Problems of Biology, Russian Academy of Sciences, Pushchino, Moscow Region, 142290, Russia}

\begin{abstract}
It is shown that the bipolaron ground state is described by a delocalized wave function. For a two-parameter wave function, the lowest variation estimate of the ground state energy in the strong coupling limit is found to be $E=- 0,414125 \alpha ^2$. This is much lower than that derived with the use of the localized bipolaron wave function. The results obtained testify to the possibility of a bipolaron mechanism of high-temperature superconductivity.
\end{abstract}


\end{frontmatter}

\section{Introduction}
\label{}
The problem of possible arising of high-temperature superconductivity (HTSC) and explanation of this phenomenon by the existence of bipolaron states was
studied in a number of papers (see reviews \cite{lit1,lit2}). There the occurrence of HTSC is treated via the mechanism of Bose-condensation of
bipolaron gas. The temperature of Bose-condensation $T_0=3,31\hbar ^2n_0^{2/3}/k_Bm_{BP}$ which in this case can be identified with the critical
temperature $T_c$ of the transition into the superconducting state at $m^*_{BP}\approx 10m_0$ depending on the bipolarons concentration $n_0$ varies in a
wide range from $T_0\approx 3K$ for $n\approx 10^{18}$ cm$^{-3}$to $T_0\approx 300K$ for $n\approx 10^{21}$ cm$^{-3}$. In the latter case the bipolarons
concentration is so considerable that for bipolaron gas as well as for superconducting pairs, a bipolaron, being a composite formation should no longer
conduct itself as an ideal Bose-particle, and at still greater concentrations it should decompose into individual polarons.  In this connection the study
of conditions at which bipolaron states are stable takes on primary importance. The lowest values of the bipolaron energy were obtained for the
electron-phonon coupling constant $\alpha < 8$  in \cite{lit3,lit4,lit5}, and for  $\alpha >8$ in \cite{lit5,lit6,lit7}.

In paper \cite{lit8} the lowest energy of the bipolaron energy was found for  $\alpha >8$. Here our aim is to improve that estimate.

\section{Results and Discussion}
\label{}

We will proceed from Froehlich Hamiltonian for the bipolaron:

\begin{eqnarray}
\label {eq.1}
\hat{H} = -\frac {\hbar ^2}{2m}\Delta _{r_1}
-\frac {\hbar ^2}{2m}\Delta _{r_2} \\ \nonumber
+\sum_k{\hbar\omega a^+_k a_k}
+U(|{\vec r}_1 - {\vec r}_2|) \\ \nonumber
+\sum_k(V_ke^{ikr_1}a_k + V_ke^{ikr_2}a_k + h.c.),\\
U(|{\vec r}_1 - {\vec r}_2|) =e^2/\varepsilon_\infty|{\vec r}_1 - {\vec r}_2|,\nonumber
\end{eqnarray}

\noindent where $m$ is the electron effective mass; $r_1$, $r_2$ are coordinates of the first and second electron, respectively; $a^+_k$, $a_k$ are
operators of birth and annihilation of phonons with energy $\hbar \omega $,

\begin{eqnarray}
\label {eq.2}
V_k = \frac {e}{k}\sqrt{\frac {2\pi\hbar\omega}{\tilde\varepsilon V}},\quad
\tilde\varepsilon ^{-1} = \varepsilon ^{-1}_{\infty} - \varepsilon^{-1}_{0},
\end{eqnarray}

\noindent where $e$ is the electron charge, $ \varepsilon _{\infty}$ and $\varepsilon_{0}$ are high-frequency and static dielectric permittivities, $V$
is the crystal's volume.

In the center-of-mass system, Hamiltonian (\ref{eq.1}) takes the form:

\begin{eqnarray}
\label {eq.3}
\hat{H} = -\frac {\hbar ^2}{2M_e}\Delta _{R}
-\frac {\hbar ^2}{2\mu _e}\Delta _{r}
+U(|{\vec r}|)
+\sum_k{\hbar\omega a^+_k a_k} \\ \nonumber
+\sum_k 2\cos\frac{\vec{k}\vec{r}}{2}(V_k\,a_k\,e^{i\vec{k}\vec{R}} + h.c.),\\ \nonumber
\vec{R} = ({\vec {r}}_1 + {\vec {r}}_2)/2,\quad
\vec{r} = ({\vec {r}}_1 - {\vec {r}}_2),\\ \nonumber
M_e = 2m,\quad \mu _e = m/2.
\end{eqnarray}

\noindent In what follows we will use the units in which $\hbar =1$, $\omega =1$, $M_e =1$ (accordingly $\mu _e=1/4$).

The center-of-mass coordinate $\vec{R} $ can be eliminated from (\ref{eq.3}) with the use of Heisenberg canonical transformation :

\begin{eqnarray}
\label {eq.4}
\hat{S}_1 = \exp {\{-i\sum_k \vec{k}\,a^+_ka_k\}\vec{R}},
\end{eqnarray}

\begin{eqnarray}
\label {eq.5}
\hat{\tilde{H}} = \hat{S}^{-1}_1\,\hat{H}\,\hat{S}_1 = -2\Delta_r
+U(|{\vec r}|)
+ \sum_k{a^+_k\, a_k}\\ \nonumber
+ \sum_k 2\cos\frac{\vec{k}\vec{r}}{2}(V_k\,a_k + V_k^*\,a^+_k)
+ \frac{1}{2}(\,\sum_k\vec{k}\,a^+_k\,a_k)^2.
\end{eqnarray}

\noindent From (\ref{eq.5}) it follows that the exact solution of the bipolaron problem, determined by the wave function $\Psi(r)$ contains only relative
coordinates $r$ and is, therefore, translationally-invariant.

Averaging $\hat{\tilde{H}}$ over $\Psi(r) $ we rewrite (\ref{eq.5}) as Hamiltonian $\hat{\bar{H}} $:
\begin{eqnarray}
\label {eq.6}
\hat{\bar{H}} =
\frac{1}{2}(\sum_k\vec{k}\,a^+_k\,a_k)^2
+ \sum_k{a^+_k\, a_k}\\ \nonumber
+ \sum_k \bar{V}_k(a_k + a^+_k)
+ {\bar T} +{\bar U},\\ \nonumber
{\bar V}_k = 2V_k\langle \Psi | \cos\frac{\vec{k}\vec{r}}{2}| \Psi\rangle,\quad
{\bar U} = \langle \Psi|U(r)|\Psi\rangle,\\ \nonumber
{\bar T} = -2\langle \Psi|\Delta _r|\Psi\rangle.
\end{eqnarray}

\noindent Lee, Low and Pines canonical transformation \cite{lit9} of this Hamiltonian:
\begin{eqnarray}
\label {eq.7}
\hat{S}_2 = \exp {\{\sum_k f_{k}\,(a_k - a^+_k)\}}
\end{eqnarray}

\noindent yields:
\begin{eqnarray}
\label {eq.8}
\hat{\tilde{\bar{H}}} = \hat{S}^{-1}_2\,\hat{\bar{H}}\,\hat{S}_2,\quad
\hat{\tilde{\bar{H}}} = \hat{H}_0 + \hat{H}_1,
\end{eqnarray}

\begin{eqnarray}
\label {eq.9}
\hat{H}_0 = {\bar T} +{\bar U} + 2\sum_k\bar{V}_kf_k
+\sum_kf^2_k + \frac{1}{2}(\sum_k\vec{k}f_k)^2 \\ \nonumber
+\sum_k(1 + \frac{k^2}{2} +\vec{k}\sum_{k^{\prime}}
\vec{k^{\prime}} f^2_{k\prime})a^+_ka_k \\ \nonumber
+\frac{1}{2}\sum_{k,\,k^{\prime}}{(\vec {k} \,\vec {k^{\prime}})} f_k\, f_{k^{\prime}} \\ \nonumber
\cdot ({a_k\,a_{k^{\prime}}+a^+_k\, a^+_{k^{\prime}} +a^+_k\, a_{k^{\prime}}+a_k \,a^+_{k^{\prime}}}),
\end{eqnarray}

\begin{eqnarray}
\label {eq.10}
\hat{H}_1 = \sum_k\lbrack\bar{V}_k + f_k
(1 + \frac{k^2}{2} +\vec{k}\sum_{k^{\prime}} \vec{k^{\prime}} f^2_{k^{\prime}})\rbrack (a_k + a^+_k)\\ \nonumber
+ \sum_{k,k^{\prime}}{(\vec {k} \,\vec {k^{\prime}})} \lbrack f_{k^{\prime}}\, a^+_k \,a_k a_{k^{\prime}} +
f_{k^{\prime}}\,a^+_k  a^+_{k^{\prime}}a_k\rbrack\\ \nonumber
+ \frac{1}{2}\sum_{k,k^{\prime}}{(\vec {k} \,\vec {k^{\prime}})a^+_k  a^+_{k^{\prime}}a_k a_{k^{\prime}}}.
\end{eqnarray}

\noindent The probe WF $\Psi_0$ for minimizing the energy determined by $\hat{\tilde{\bar{H}}}$ is chosen as:
\begin{eqnarray}
\label {eq.11}
\Psi _0 = \Psi (r)\exp \{-i\sum _k\vec k a^+_k a_k \vec R\}
\\ \nonumber
\cdot \exp \{ \sum _k f_k(a_k - a^+_k)\}\Lambda _0,
\end{eqnarray}

\noindent where $\Lambda _0$ is the eigen wave function of operator (\ref{eq.9}) \cite{lit8,lit10}. As a result we express the energy $E$ of the
bipolaron ground state as:
\begin{eqnarray}
\label {eq.12}
E = \Delta E
+ 2\sum_k \bar{V}_k f_k
+ \sum_k{f^2_k}
+{\bar T} +{\bar U},
\end{eqnarray}

where
\begin{eqnarray}
\label {eq.13}
\Delta E = \frac{1}{4\pi ^2}
\int^{\infty}_0{\frac{k^4f^2_kdk}{(1 + Q)}}
+ \frac{1}{12\pi ^4}
\int^{\infty}_0 k^4p^4f^2_kf^2_p
\\ \nonumber
\cdot
{\frac{\omega _p(\omega _k\omega _p + \omega _k(\omega _k + \omega _p) + 1)}{(\omega _k + \omega _p)^2(\omega _p^2 - 1)|D_+(\omega _p^2)|^2}}dpdk ,\\ \nonumber
Q = \frac{1}{3\pi ^2}\int^{\infty}_0{\frac{k^4f^2_k\omega _k}{(\omega _k^2 - 1)}}dk.
\end{eqnarray}

\noindent Notice that in accordance with \cite{lit8,lit10} the bipolaron wave function is delocalized with respect to R.
To get the energy of bipolaron ground state we will use variational approach and assume:

\begin{eqnarray}
\label {eq.14}
f_k=-\bar{V}_k\exp (-k^2/2\mu),\\ \nonumber
  \Psi(r)=(\frac{2}{\pi l^2})^{3/4}\exp (-\frac{r^2}{l^2})
\end{eqnarray}
where $\mu$ and $l$ are variational parameters. Notice, that (\ref{eq.14}) transforms to the results of paper \cite{lit8} in the limit
$\mu\rightarrow\infty$. With the use of (\ref{eq.14}) we express $E$ as:

\begin{eqnarray}
\label {eq.15}
E = \frac {20,25}{l^2+16/\mu}-\frac {32\alpha }{\sqrt\pi} \frac {1}{\sqrt{l^2+8/\mu}}\\ \nonumber
+\frac {16\alpha }{\sqrt\pi} \frac {1}{\sqrt{l^2+16/\mu}}+\frac {6}{l^2}\\ \nonumber
+\frac {4\sqrt{2}}{\sqrt\pi}\frac {\alpha }{1-\eta} \frac {1}{l},
\end{eqnarray}

\noindent where $\eta=\varepsilon_{\infty} /\varepsilon_0$ $\alpha=(e^2/\hbar \tilde{\varepsilon})\sqrt{m/2\hbar\omega}$ - is a constant of
electron-phonon coupling.

Let us write down $E_{min}(\eta)$ for the minimum of the bipolaron total energy (15) with respect to the parameters $\mu$ and $l$. Fig. (1) shows the
dependence of $E_{min}(\eta)$  (fig. 2 demonstrates the dependencies of $l_{min}$, $\mu _{min}$ on $\eta$) on the parameter $\eta$ in the range of its
variation (0, 1). Fig. 1 suggests that $E_{min}(0)=-0,414125 \alpha ^2$. Accordingly, for the critical value of the ion coupling parameter at which the
bipolaron state is impossible it is equal to $\eta _{kp}=0,317$, $E_{min}(\eta _{kp})=2E_p$ where $E_p=-0,1085128 \alpha ^2$ is the polaron energy
\cite{lit11}. In \cite{lit8} for $\eta _{kp}$, was found to be: $\eta _{kp}=0,2496$ . The results obtained lead us to a considerably lower threshold of
the formation of bipolaron states with respect to the electron-phonon interaction constant: $\alpha _c\geq4,829$ which significantly expands the range of
crystals in which bipolarons are possible. Previously the lowest value of $\alpha _c=6,8$ was obtained in \cite{lit4}.

\begin{figure}[t]%
\includegraphics*[width=\linewidth,height=\linewidth]{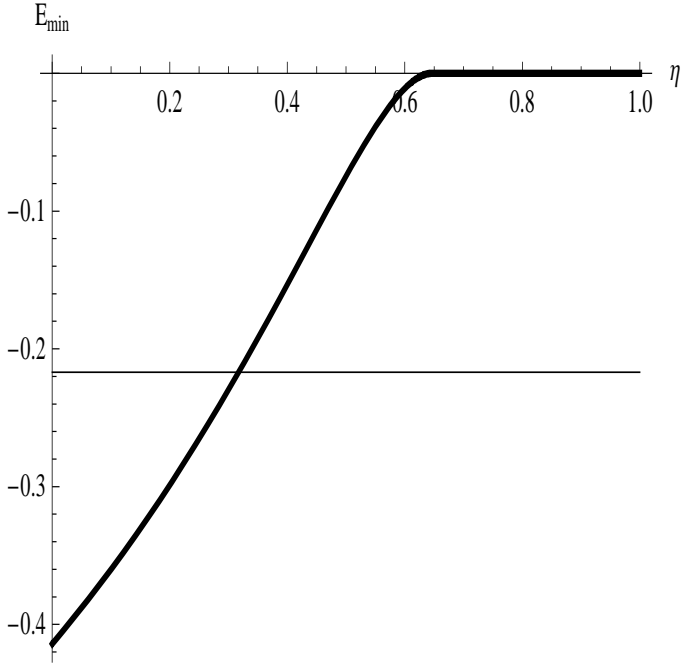}
\caption{%
  Function $E_{min}(\eta)={min}_{x,y} E(x,y,\eta)$; a line $E=-0.217; E_{min}(0)=-0.414125$.}
\label{onefig}
\end{figure}

\begin{figure}[ht]%
\includegraphics*[width=\linewidth,height=\linewidth]{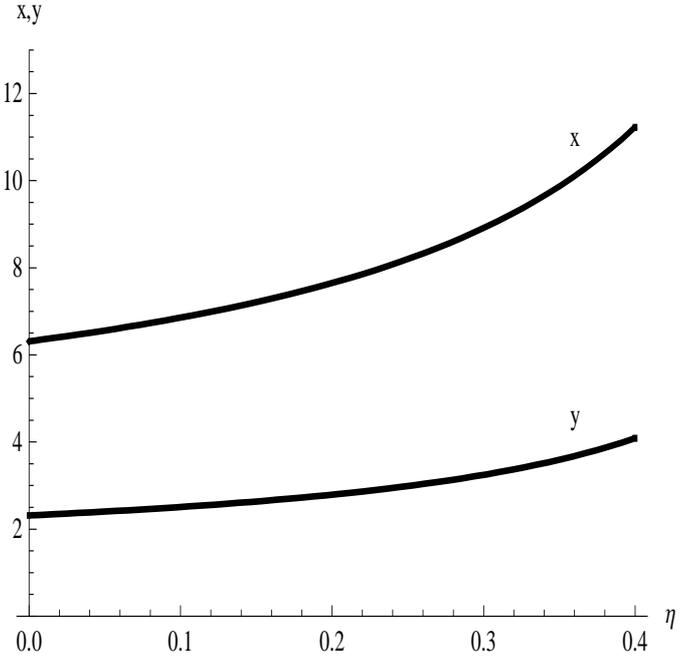}
\caption{%
  The dependence of $x=x_{min}(\eta)$, $y=y_{min}(\eta)$on $\eta $; $x_{min}(\eta)$, $y_{min}(\eta)$ correspond to minimum of $E(x.y;\eta)$, $x=\alpha l$, $y=\alpha ^2/\mu$.}
\label{twofig}
\end{figure}

Notice that $E_{min}$ determined by expression (\ref{eq.14}) yields much lower values of the bipolaron energy than all the results obtained earlier for all the values of $\eta$, and, in particular, the value of $E_{min}(0)=-0,3243 \alpha ^2$($\eta _{kp}=0,2496$) found in \cite{lit8} and $E_{min}(0)=-0,243628 \alpha ^2-4,22606 +0(\alpha ^{-2})$ derived in \cite{lit12}.

According to (\ref{eq.14}), the characteristic size of a bipolaron state is equal to $l$ and in dimensional units is: $l=\hbar ^2\tilde{\varepsilon}x/me^2$, where the dependence $x(\eta)$ is determined by that given in fig.2. From fig. 2 follows that on the whole interval of $\eta$ variation in the range of bipolaron states stability $\eta\in $($0;\eta _c$), the value $x$ changes only slightly from $x(\eta =0)\approx6,3$ to $x(\eta =0,317)\approx9,2$.  Hence, even for $\eta=\eta _c$, the critical concentration of bipolarons at which their composite character makes itself evident, is $\eta _c\cong10^{21}$cm $^{-3}$. If we take the maximum value of  $\eta _c$ for a high-temperature superconductor $La_2CuO_4$ to be equal to $\eta _c=\eta _{c\bot}=0,174$ (the value of $\eta$ in the direction perpendicular to the $CuO_2$ layers \cite{lit13}), then the critical value of the concentration will be an order of magnitude greater than $\eta _c=10^{21}$cm$^{-3}$. These results provide support for the conclusion that 3D bipolaron mechanism of superconductivity is possible in copper oxides.

Being delocalized, these bipolaron states are translation-invariant (TI) and for $P=0$, where $P$ is the total momentum of the system, are separated by an energy gap from bipolaron states with spontaneously broken translation symmetry described by localized wave function \cite{lit2}. For $P\neq0$, TI bipolarons are not scattered on optical (acoustical) oscillations of the lattice \cite{lit10}, besides, as delocalized, they are not scattered on local defects of the lattice either.
As is known \cite{lit14,lit15,lit16}, explanation of high-temperature superconductivity by Bose-concentration of bipolarons faces some problems associated with a large mass of these formations and, as a consequence, low values of $T_c$. The above indicated features  of translation invariant bipolarons confer them superconducting properties even if their Bose-condensation is absent.

The work was supported by RFBR projects N 11-07-12054; 10-07-00112.

\bibliographystyle{elsarticle-num}
\bibliography{lakhno}

\end{document}